# Machine Learning-based Test Selection for Simulation-based Testing of Self-driving Cars Software

Sajad Khatiri, Christian Birchler,
Bill Bosshard, Alessio Gambi,
Sebastiano Panichella

Abstract Simulation platforms facilitate the development of emerging cyber-physical systems (CPS) like self-driving cars (SDC) because they are more e -cient and less dangerous than eld operational tests. Despite this, thoroughly testing SDCs in simulated environments remains challenging because SDCs must be tested in a sheer amount of long-running test scenarios. Past results on software testing optimization have shown that not all the tests contribute equally to establishing con dence in test subjects' quality and reliability, with some \uninformative" tests that can be skipped (or removed) to reduce testing e ort. However, this problem was partially addressed in the context of SDC simulation platforms. In this paper, we investigate test selection strategies to increase the cost-e ectiveness of simulation-based testing in the context of SDCs. We propose an approach called SDC-Scissor (SDC coSt-e eCtIve teSt SelectOR), which leverages machine learning (ML) strategies to iden-tify and skip tests that are unlikely to detect faults in SDCs before executing them. Speci cally, SDC-Scissor extract features concerning the characteristics of the test scenarios being executed in the simulation environment and via

Sajad Khatiri
Zurich University of Applied Science & Software Institute - USI, Lugano, Switzerland E-mail: mazr@zhaw.ch, mazras@usi.ch

Christian Birchler and Sebastiano Panichella
Zurich University of Applied Science, Switzerland
E-mail: birc@zhaw.ch, panc@zhaw.ch

Bill Bosshard
Meier Planungsdienste GmbH, Switzerland.
E-mail: bill.bosshard@outlook.com

Alessio Gambi
University of Passau, Germany.
E-mail: Alessio.gambi@uni-passau.de

ML strategies predict tests that lead to faults before executing them. Our evaluation shows that SDC-Scissor achieved high classi cation accuracy (up to 93.4%) in classify tests leading to fault which allows to improve testing cost-e ectiveness: SDC-Scissor was able to reduce (ca. 170%) the time spent in running irrelevant tests as well as identi ed 33% more failure triggering tests compared to a randomized baseline. Interestingly, SDC-Scissor does not intro-duce signi cant computational overhead in the SDCs testing process, which is critical to SDC development in industrial settings.

Keywords Self-driving cars, Software Simulation, Regression Testing, Test Case Selection.

## 1 Introduction

Cyber-physical systems (CPSs) leverage physical capabilities from hardware components as well as computational and arti cial intelligence from software components to operate in complex and dynamic environments, potentially in-volving humans [12]. Speci cally, CPSs collect, analyze, and leverage sensor data from the surrounding environment continuously to control physical actua-tors at run-time [4,12]. CPSs nd application in various domains ranging from Robotics and Transportation to Healthcare and are expected to drastically improve the quality of life of citizens and the economy [20].

Among various and emerging CPS application domains, the usage of self-driving cars (SDCs) in transportation is expected to impact our society pro-foundly. Human errors cause more than 90% of driving accidents (e.g., driving while under the in uence of alcohol, fatigue, and other distractions) [39]; hence, automated driving systems such as SDCs have the potential to reduce such errors and eliminate most accidents. However, the recent fatal crashes involv-ing self-driving cars suggest that the advertised large-scale adoption of SDCs appears optimistic and premature [12,34]. One of the main factors limiting the usage of autonomous driving solutions is the lack of adequate testing. Con-sequently, the risk of releasing SDCs equipped with defective software, which might become erratic and lead to fatal crashes, is still quite high [34].

Testing automation is crucial for ensuring the safety and reliability of SDCs [39, 42]. However, most developers rely on human-written test cases (at unit and system levels) to assess SDCs' behavior. This practice has sev-eral limitations and drawbacks: (i) limited possibility to repeat tests under the same conditions [42]; (ii) di culty in testing SDCs in representative and safety-critical scenarios [34, 37, 59]; (iii) di culty in assessing SDC's behavior in di erent environments and execution conditions [39].

As a consequence, SDCs practitioners in the eld are facing a fundamental development challenge: observability, testability, and predictability of the be-havior of SDCs are highly limited [34, 37, 59]. Thus, new testing practices and tools are needed to nd SDC faults earlier during development and, eventually, support the widespread usage of autonomous driving.

The utilization of simulation environments can potentially address several of the challenges mentioned above [1, 13, 15, 25] since simulation-based testing is more efficient than and can be as effective as traditional field operational testing [5,25]. Additionally, simulation-based testing can support and complement well-established hardware-in-the-loop (HiL), model-in-the-loop (MiL), and software-in-the-loop (SiL) development strategies. Consequently, an in-creasingly large number of commercial and open-source simulation environ-ments have been delivered to the market to conduct testing in the autonomous driving domain [13, 25] as well as other CPS domains [55].

## 1.1 Problem Statement and Summary of Results

The usage of simulation environments enables automated test generation and execution [31]. However, the size of the testing space of simulation environ-ments is in principle infinite, which poses the main open challenge of exercising the SDC behaviors adequately [3,31]. The budget devoted to testing activities is usually limited, making the identification of faults particularly challenging in the SDC domain since the execution of simulation-based tests is considerably slower compared to other forms of tests (e.g., unit tests) [27, 63]. Therefore, it is paramount that developers test SDCs cost-effectively, for example, by using test suites optimized to reduce testing effort without affecting their ability to identify faults in SDCs using simulations both in nominal operating conditions and corner cases [3, 48, 64].

In this paper, we investigate test case selection (TCS) techniques to im-prove the cost-effectiveness of simulation-based testing in the context of SDCs. Specifically, we focus on techniques that employ Machine Learning (ML) mod-els to optimize the TCS cost-effectiveness. The main challenges we focus on while designing such ML-based test case selection strategies for SDCs are as follow: (i) the definition of the features that best characterize faulty and non-faulty SDC test scenarios; (ii) the identification of suitable ML models that can reliably predict the SDCs' behavior before executing those test scenarios; and (iii) the usage of ML strategies to effectively distinguish relevant (faulty) from irrelevant (non-faulty) test scenarios.

We are interested in testing the safety of SDCs; therefore, we deem as relevant those scenarios that expose a fault (e.g., a SDC drives out of the road). We call those scenarios unsafe. Consequently, our TCS techniques exploit ML models to classify SDC test scenarios that are unsafe (i.e., likely to expose a fault) or safe.

In this paper, we seek to answer the following research questions:

RQ1: To what extent is it possible to identify safe and unsafe test scenarios for SDCs before executing them?

We focus on designing driving scenarios input features, i.e., features that concern the SDC simulation-based tests and can be extracted before their execution. Thus, we propose SDC-Scissor, a framework the leverages the aforementioned features to train machine-learning models that classify test scenarios as safe or unsafe. Specically, to distinguish between safe and un-safe test scenarios we focus on lane-keeping functionalities in which unsafe scenarios cause a self-driving car to depart its lane [31] and investigate features that either describe the geometry of a road as a whole (i.e., full road) or describe properties of the road segments comprising it. Finally, we investigate the accuracy of SDC-Scissor in classifying safe and unsafe test scenarios of SDCs.

RQ2: Does SDC-Scissor improve the cost-e ectiveness of simulation based testing of SDCs?

We investigated whether SDC-Scissor reduces testing time dedicated to execute irrelevant (safe) tests while keeping a high test cost-e ectiveness (i.e., identify th same of higher number of safe tests without impacting test costs). We study SDC-Scissor's behavior in two opposite setups and con-textualize our ndings by comparing the results against a random baseline approach (i.e., the scenarios are randomly generated, selected, and exe-cuted). In the rst study, SDC-Scissor leverages ML models trained on o -line data (i.e., trained on a large static dataset); this setup lets us eval-uate the application of the proposed technique for regression testing. In the second study, SDC-Scissor leverages real-time data (i.e., dynamically gen-erated tests) and continuously (re-)trained ML models; this setup lets us evaluate the application of the proposed technique for automated test gen-eration. As described before, in both setups we compared the time-saving ability of SDC-Scissor with respect to the random selection strategy as well as its ability to detect more faults while allocating lower test execu-tion costs.

We conducted our investigation using the freely available SDCs simula-tor BeamNG.tech [13] (elaborated in Section 2) and the open-source tool AsFault [31]. We selected BeamNG.tech because it can execute procedurally generated driving scenarios, and it was recently adopted as the reference sim-ulator in the ninth edition of the Search-Based Software Testing tool compe-tition [1] [50]. We selected AsFault because it can automatically generate test scenarios to assess SDCs' lane-keeping and is compatible with BeamNG.tech. It is important to note that in the rest of the paper we will refer to test scenarios generated with AsFault as test cases, to avoid any confusion in terminology.

Our results show that SDC-Scissor achieved high prediction accuracy (be-tween 72% and 96% in predicting unsafe test cases). Not only SDC-Scissor avoided the execution of 50% of unnecessary tests as well as identi ed 35% more unsafe test cases compared to the random baseline approach.

Our assessment of SDC-Scissor shows that SDC-Scissor successfully selects test cases independently from the AI engine used or di erent risk levels (i.e., di erent driving styles), with the Logistic model providing the more stable results. Interestingly, our results also show that the knowledge is not trans-ferable from one AI engine to another one, i.e., SDC-Scissor performed worse when training ML models on data from a specic AI engine and testing on data from a dierent AI engine. Finally, SDC-Scissor does not introduce signicant computational overhead in the SDCs testing process, critical to SDC development in industrial settings.

---

[1] https://sbst21.github.io/tools/

## 1.2 Paper contributions

The contributions of this paper can be summarized as follows:

1. Feature sets: We qualitatively and quantitatively investigated essential input features that can be used to characterize safe and unsafe test cases before executing them.
2. Selection of SDCs test cases: We investigated new methods in the area of SDCs for test case selection. Hence, we introduced SDC-Scissor that leverages ML models to improve testing cost-eectiveness via test case selection.
3. Offine v.s. Real-time Training: We investigated two opposite setups for SDC test case selection that leverage ML models trained on o-line data (i.e., trained on a large static dataset) and real-time data (i.e., dynamically generated tests).
4. Replication package: We built a large dataset of labeled test cases that can be used for replication purposes and future research [41].

## 1.3 Paper structure

The paper proceeds as follows: Section 2 provides some background about regression testing and automated test generation in the context of SDCs and CPSs. Section 3 describes the empirical study design, while Section 4 presents its main ndings. Section 5 discusses related work, while Section 6 discusses the threats that could aect the validity of our results. Finally, Section 7 concludes the paper and outlines future research directions.

## 2 Background on Regression Testing and Simulation for CPSs

This section (i) briey discusses the test optimization approaches in traditional systems and reects upon existing test selection strategies in the context of CPSs; then, (ii) introduces background elements to make this paper self-contained.

## 2.1 Software Testing Optimization

Research has yielded many approaches to optimize testing. However, most of the available approaches focus on regression testing and found application only in traditional software systems [65]. These approaches can be classi
ed into the following categories: test case selection [21], test suite, and test
case minimization) [53], and test case prioritization [54].

Test case selection approaches identify subsets of available tests relevant
(or necessary) for testing a given change in the code. Test suite reduction
approaches remove redundant test cases from existing test suites leading to
smaller test suites that can execute faster, while test case minimization ap-
proaches remove irrelevant statements from the tests, reducing their size. Fi-
nally, test case prioritization approaches rank test cases by the likelihood of
detecting faults such that their execution can lead to nding faults soon.

Compared to traditional software systems, CPSs face additional challenges
due to their continuous interactions with the environment and the tight cou-pling
between the hardware and software components comprising them. Conse-
quently, when it comes to testing, standard testing approaches are ine ective,
ine cient, or inapplicable [16]. Testing of CPSs has typically been performed
following X-in-the-loop paradigms [46] that in practice takes the form of the
model in the loop (MiL), software in the loop (SiL), and hardware in the loop
(HiL). For MiL testing, most of the software and hardware components, sen-
sors, and other relevant environmental elements are abstracted using models,
such that testing can focus on assessing the correctness of the control algo-
rithms governing the CPSs. For SiL testing, only the hardware components and
the environments are abstracted using physically accurate simulators; in this
case, the behavior and integration of CPSs software components can be tested
in realistic execution conditions. Finally, HiL testing focus on checking the
integration of hardware and software components in production-like, ei-ther
simulated or real, environments. Notably, X-in-the-loop testing involves a great
deal of simulation.

Considering the speci c need for X-in-the-loop development of CPSs, re-
searchers proposed testing optimization techniques tailored for CPSs [7{10,56].
Test selection (or prioritization) for traditional systems is typically performed by
computing the test similarity or test adequacy (i.e., code coverage). How-ever,
given the complexity of test inputs for CPSs (e.g., simulated environ-ments),
computing traditional similarity metrics based on lexicographic sim-ilarity of test
code and test inputs is technically challenging and may not be adequate.
Consequently, new similarity metrics and procedures to compute them have
been proposed recently. For instance, Arrieta et al. [7, 9] proposed to measure
the similarity between the test cases based on the so-called signal values of all
the states for the simulation-based test cases.

Adopting code coverage as a proxy for test adequacy in CPSs systems,
which are based on arti cial intelligence and deep learning, is not adequate;
hence, selecting tests purely driven by code coverage is bound to produce ine
ective test suites. Because of this, current research e orts focus on di erent
heuristics to select test cases. Arrieta et al. [8] proposed to select test guided by
objectives such as requirement coverage and test execution times; they applied
test selection in the context of multi-objective test case generation for CPSs and
showed improvement over baseline approaches. Complementary, Shin et al. [56] proposed an approach for acceptance test selection for a satellite system that selects relevant test cases based on two objectives: (i) the risks of causing hardware damage, and (ii) the number of test cases executed within a given time budget.

Compared to those studies, we investigate (1) a different CPS domain and (2) different test selection objectives. Specifically, we investigate lane-keeping in self-driving car simulation environments, whereas previous work focused on industrial tanks [8,10], satellites [55], electric windows [7] and cruise controllers [8]. Regarding test selection objectives, we focus on improving the cost-effectiveness of simulation-based tests to assess safety requirements. In contrast, previous studies prioritized the execution of tests based on their fault-detection capability [10], or selected tests based on signals diversity [7{9], that require at least one execution of the test cases in both approaches. Since in the SDC domain, executing simulation-based tests is prohibitive, we face the challenge of selecting test cases before their execution. Consequently, our techniques consider only the initial state of the car and the characteristics of the roads (e.g., geometry, lane markings), as those features are available without executing the tests in the simulator.

## 2.2 Background on CPS Simulation

Several simulation technologies have been developed to support developers in various stages of the design and validation of CPSs. For instance, in the self-driving cars domain, developers resort to basic simulation models [33, 58], rigid-body simulations [45,67], and soft-body simulations [29,52] among others.

Basic simulation models, like MATLAB and Simulink models, have been mainly utilized for model-in-the-loop simulations and Hardware/Software co-design. They implement fundamental abstractions (e.g., signals) but target mostly non-real-time executions and generally lack photo-realism; consequently, their usage as a means for testing lane-keeping and other vision-based systems is limited.

Rigid-body simulations approximate the physics of bodies by modeling entities as undeformable bodies or as compositions of a limited number of rigid three-dimensional objects such as boxes, cylinders, and convex meshes [3]. Rigid-body simulations implement a coarse approximation of reality; hence, they can efficiently simulate basic object motions and rotations and scale well in the number of simulated entities (e.g., vehicles). However, they can simulate breaks only inaccurately and cannot simulate body deformation at all.

Soft-body simulations, instead, can simulate deformable and breakable objects and fluids; hence, they can handle a wide range of simulation cases in addition to primitive body motions and rotations. Mass-spring systems and finite element method (FEM) are the main approaches to simulate solid objects, while finite volume method (FVM) and finite difference method (FDM) are the main approaches for simulating fluids [47]. For simulating SDCs, mass-spring systems and FEM are the most suited soft-body approaches since they target solid objects. Both approaches model solid objects as a composition of (many) atomic elements interacting with each other and reacting to external forces. Therefore, the simulations follow a bottom-up approach: the high-level behavior of the simulated objects emerges from the simulation of the behavior of the atomic elements comprising them.

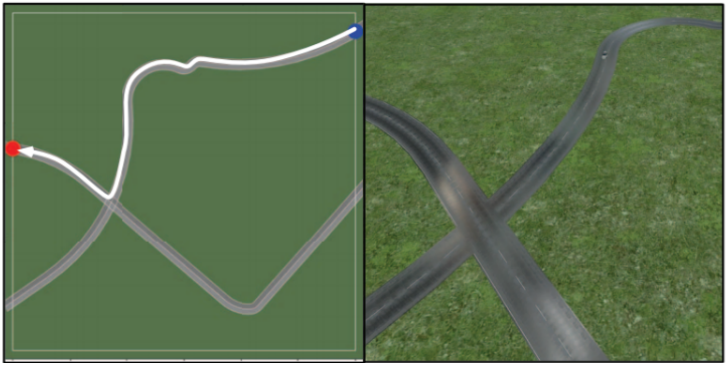

Fig. 1 Example for simple test case by AsFault [31]

Rigid-body v.s. Soft-body SDC simulations. Both rigid- and soft-body simulations can be effectively combined with powerful rendering engines to implement photo-realistic simulations [13, 15, 25, 62]; consequently, both approaches are viable solutions for simulating SDCs. However, soft-body simulations can simulate a wider variety of physical phenomena compared to rigid-body simulations. For example, soft-body simulations can model body deformations, fractures, vibrations, anisotropic mass distributions, and inertia, essential in many CPSs scenarios. Soft-body simulations are also very versatile. As stated by Dalboni and Soldati [24] speaking about mass-spring systems, using the elementary description of target systems as collections of nodes and beams (i.e., springs), it is possible to contemplate all the laws of mechanics that rule the physical world. Consequently, soft-body simulations can simulate different materials and other phenomena, such as aerodynamics and pressured volume changes relevant in many CPSs domains. Soft-body simulations are more accurate than rigid-body simulations; hence, they are more computationally demanding and less scalable in the number of simulated bodies. Consequently, soft-body simulations are less suitable than rigid-body simulations for simulating complicated traffic scenarios where the movement of the simulated entities is generally unrestricted (i.e., there are no collisions between the simulated entities). In contrast, soft-body simulations are a better fit for implementing safety-critical scenarios (e.g., car crashes [29]) and focused scenarios in which high simulation accuracy, even in extreme situations, matters the most (e.g., simulating an unbalanced load of trucks or driving with a flat tire).

## 2.3 SDC Test Case Generation Environment

Manually creating adequate test scenario suits for SDCs is a complex and laborious task as it requires testers to have experience in multiple domains, including CPS, simulations, physics, and 3D object modeling. In order to tackle this issue, Gambi et al. [32] proposed a search-based approach for procedurally generating driving scenarios (as in Figure 1) for testing lane-keeping systems. This approach is implemented in an open-source tool called AsFault [31]. This section brie y summarizes how AsFault generates simulation-based tests, as we will use it to evaluate our ML-based test selection techniques.

AsFault generates virtual roads for testing lane-keeping systems and leverages a genetic algorithm to re ne those virtual roads until they become so challenging for the system under test (i.e., the lane-keeping system driving the ego-car) that they cause the ego-car to drive o the lane. When this hap-pens, we say that AsFault identi ed an Out of Bound Episodes (OBEs), i.e., a safety-critical issue. AsFault reports the total count of OBEs and the road segments where OBEs have been observed for each generated scenario. In our experiments, we use this information in order to label test scenarios as safe (causing no OBEs) or unsafe (causing at least an OBE). AsFault relies on BeamNG.tech [13] for executing the generated tests as physically accurate and photo-realistic driving simulations (see Figure 1).

We consider two lane-keeping systems as test subjects for our evaluation: The rst, BeamNG.AI[2], is the driving agent shipped with the BeamNG.tech, and the second, Driver.AI[3], is a trajectory planner shipped with AsFault. These test subjects have perfect knowledge of the virtual roads and drive the ego-car by computing an ideal driving trajectory to stay in the center of the lane while driving within a con gurable speed limit. As explained by BeamNG.tech developers, a parameter called the \aggression" factor controls the driving style of BeamNG.AI: low aggression values (e.g., 0.7) result in smooth driving, whereas high aggression values (e.g., 1.2 and above) result in an edgy driving that may lead the ego-car to \cut corners". Driver.AI instead analyzes the road geometry and plans the car trajectory by computing for each turn the maximum safe driving speed (v) using the standard formula for centripetal force on at roads with static friction ( ) [22]:

$$v = \sqrt{r\ g}$$
(1)

where r is the turn radius and g is the free-fall acceleration. Driver.AI relies on the user to provide the value of the friction coe cient, as well informa-tion about the maximum acceleration and deceleration of the ego-car. In our evaluation, we estimated those values empirically following a trial-and-error approach.

---

2  https://wiki.beamng.com/Enabling_AI_Controlled_Vehicles#AI_Modes

3  https://github.com/alessiogambi/AsFault/blob/asfault-deap/src/asfault/ drivers.py

Table 1 Full Road Attributes. In the table, we report for each feature their name, descrip-tion, type, and range. We computed the range empirically

| Feature | Description | Range |
|---|---|---|
| Direct Distance | Euclidean distance between start and finish (Meters) | [0 { 489.9] |
| Length | Total length of the driving path (Meters) | [50.6{3317.9] |
| Num L Turns | Number of left turns on the driving path | [0 { 18] |
| Num R Turns | Number of right turns on the driving path | [0 { 17] |
| Num Straight | Number of straight segments on the driving path | [0 { 11] |
| Total Angle | Cumulative turn angle on the driving path | [105{ 6420] |

## 3 Study Design

In this paper, we investigate Machine Learning-based test selection techniques for improving the cost-e ectiveness of simulation-based testing for SDCs. The rst challenge concerns the identi cation of features that can be used to predict whether test cases are safe or unsafe (RQ$_1$). We focus on extracting features from test case de nitions (e.g., road features and road segment features), i.e., features available before executing the simulations. The second challenge is de-vising techniques that e ectively leverage such test input features to minimize testing costs while keeping testing e ectiveness high (RQ$_2$). Speci cally, we investigate two alternative setups (explained later in this section): pre-trained ML models (referred to o ine training later), which may nd application in regression testing, and real-time retrained ML models, which are suitable in automated test generation.

This section describes the design of our empirical study, including the preparation of the training and testing datasets, the adopted research method, and the experimental settings. The following section, instead, elaborates on the achieved experimental results.

### 3.1 Dataset Preparation

As discussed in Section 2, we used AsFault to generate the test cases that form our dataset. AsFault provides a set of attributes that can describe some aspects of the generated test cases. We consider those attributes as potential input features and re ne them as described in the remainder of this section. We also used AsFault for executing the test cases to obtain the required labels (safe or unsafe) to train the ML models.

#### 3.1.1 SDC Test Case Feature Sets and Labeling

To predict whether test simulations likely result in safe or unsafe test cases before their execution, we designed two sets of input features: Full Road and Road Segment features. The former set of features concerns global attributes of the virtual roads used as test cases, while the latter focuses on the local characteristics of the road segments forming the virtual roads. The goal is to understand whether ML models trained using global features have the same prediction power as ML models trained using local features or not.

Table 2 Full Road Statistics. In the table, we report for each feature their name, descrip-tion, type, and range. We computed the range empirically

| Feature | Description | Range |
|---|---|---|
| Median Angle | Median turn angle on the driving path | [30 {330] |
| Std Angle | Standard deviation of turn angles on the driving path | [0 {150] |
| Max Angle | Maximum turn angle on the driving path | [60 {345] |
| Min Angle | Minimum turn angle on the driving path | [15 {285] |
| Mean Angle | Average turn angle on the driving path | [52.5[307.5] |
| Median Radius | Median turn radius on the driving path | [7 {47] |
| Std Radius | Standard deviation of turn radius on the driving path | [0 {22.5] |
| Max Radius | Maximum turn radius on the driving path | [7 {47] |
| Min Radius | Minimum turn radius on the driving path | [2 {47] |
| Mean Radius | Average turn radius on the driving path | [5.3 {47] |

- Full Road Features describe global characteristics of SDC test cases, such as the total length of the virtual road, its starting and target positions on the map, and the count of left and right turns. We extract two types of full road features describing the main road attributes (see Table 1) and some statistics about the road composition (see 2). We calculate road statistics in three steps. First, we extract the driving path that the ego-car must follow during the test execution; this path de nes the test case and contains the road segments that the ego-car must traverse to reach the target position from the starting position. Next, we extract the available metrics from each road segment (i.e., length for straight road segments and road angle and pivot radius for turns). Finally, we compute the statistics by applying standard aggregation functions (e.g., minimum, maximum, average) on the collected road segments metrics.

- Road Segment Features describe particular characteristics of the road seg-ments that compose a test case (see Table 3). Given the path that the ego-car must follow, we determine features that describe single road segments (e.g., is this the rst segment in the path or the last one? ) and features that correlate adjacent road segments (e.g., is the segment before this one a left turn? is the segment after this one a sharper turn? )

For each considered test case, we extract one full road data point and multiple road segment data points[4] and label them as unsafe, if AsFault reports an OBE during the simulation, or safe otherwise.

---

[4] The actual number of road segment data points depends on how many segments compose the scenario's driving path.

Table 3 Road Segment Features. In the table, we report for each feature their name, description, type, and range. We computed the range empirically

| Feature | Description | Range |
|---|---|---|
| First | Is this the rst segment on the road? | T / F |
| Last | Is this the nal segment on the road? | T / F |
| Right Turn | Is the segment a right turn? | T / F |
| Left Turn | Is the segment a left turn? | T / F |
| Straight | Is the segment a straight? | T / F |
| Angle | The angle of the segment (if turn) | [ 120 { 120] |
| Radius | The radius of the segment (if turn) | [0 { 47] |
| Length | Length of the segment | [2.1 { 393.6] |
| Direct Distance | Euclidean distance between segment's endpoints | [2.1 { 325.6] |
| Prev Right Turn | Is the previous segment a right turn? | T / F |
| Prev Left Turn | Is the previous segment a left turn? | T / F |
| Prev Straight Seg | Is the previous segment a straight segment? | T / F |
| Prev Angle | The angle of the previous segment (if any) | [ 120 { 120] |
| Prev Radius | The radius of the previous segment (if any) | [0 { 47] |
| Prev Length | Length of the previous segment (if any) | [0 { 393.6] |
| Prev Direct Distance | Euclidean distance between segment's endpoints (if any) | [0 { 325.6] |
| Next Right Turn | Is the next segment a right turn? | T / F |
| Next Left Turn | Is the next segment a left turn? | T / F |
| Next Straight Seg | Is the next segment a straight? | T / F |
| Next Angle | The angle of the next segment (if any) | [ 120 { 120] |
| Next Pivot O | The radius of the next segment (if any) | [0 { 47] |
| Next Length | Length of the next segment (if any) | [0 { 393.6] |
| Next Direct Distance | Euclidean distance between segment's endpoints (if any) | [0 { 325.6] |

## 3.1.2 Test Scenario Dataset

To achieve a comprehensive set of test cases, we considered many randomly generated test cases and test subjects in multiple con gurations (Section 2.3). We considered random tests to have an unbiased sampling of the test space and multiple test subjects and con gurations to draw conclusions about the generalizability of the proposed techniques. As reported in Table 4, we generated 8; 500 test cases and collected labels from 14; 100 simulations; in total, we collected approximately 163; 000 road and road segment data points. This section brie y summarizes the process we followed to build this dataset.

As described in Section 2, BeamNG.AI's driving style can be in uenced by setting its aggression factor (AF). Therefore, we considered three AF val-ues ranging from cautious (AF 1.0) to moderate (AF 1.5) to reckless (AF 2.0). Using di erent values for the aggression factor enables us to study the e ectiveness of our techniques concerning various SDCs' driving styles. To study the generality of our techniques, instead, we consider a second test sub-ject, Driver.AI. Speci cally, we tested Driver.AI with the same test cases used for testing BeamNG.AI in the moderate con guration. This way, we can di-rectly compare the results achieved by both test subjects. From the data in Table 4, we make the following two observations. First, the number of un-safe tests increased with increasingly large values of BeamNG.AI's aggression factor. Second, testing Driver.AI resulted in fewer unsafe cases than testing BeamNG.AI in the moderate configuration. The above observations suggest that the aggression factor strongly influences the safety of BeamNG.AI; hence, changing its value likely results in different driving styles. At the same time, Driver.AI drives more cautiously than BeamNG.AI in the moderate configuration. Therefore, different test subjects indeed drive differently on the same roads.

Table 4 Dataset Summary

| Test Subject | Feature Set | Data Points | | |
|---|---|---|---|---|
| | | Unsafe | Safe | Total |
| BeamNG.AI cautious | Full Road | 312 (26%) | 866 (74%) | 1'178 |
| BeamNG.AI moderate | Full Road | 2'543 (45%) | 3'095 (55%) | 5'638 |
| BeamNG.AI reckless | Full Road | 1'655 (96%) | 74 (4%) | 1'729 |
| Driver.AI | Full Road | 1'045 (19%) | 4'585 (81%) | 5'630 |
| BeamNG.AI moderate | Road Segment | 2'543 (3%) | 72'433 (97%) | 74'976 |
| Driver.AI | Road Segment | 2'494 (3%) | 71'145 (97%) | 73'639 |

## 3.2 Research Method

We designed three experiments to answer our research questions: The first set of experiments (i.e., Machine Learning-based Experiments) investigates whether ML models trained with global and local road features can identify safe and unsafe test cases before their execution (RQ$_1$). The second and third set of experiments (i.e., Offline Experiments and Real-Time Experiments) investigate if and how much SDC-Scissor improves the cost-effectiveness of SDC simulation-based testing (RQ$_2$).

### 3.2.1 Machine Learning-based Experiments

We study whether ML models can predict if a scenario is safe or unsafe and which combinations of features allows to achieve the more accurate prediction results. Therefore, we train various ML models and classify the test cases generated by AsFault while testing BeamNG.AI and Driver.AI.

We used Weka [28] to train and evaluate standard ML models that have been successfully used for defect prediction in software engineering in the past (e.g., [14, 19, 40]):

- Logistic Regression that uses a logistic function to model the probability of observing a certain class [60].
- J48 that creates a decision tree following the well-known C4.5 algorithm [28].
- Random Forest that uses an ensemble of decision trees [36].
- Naive Bayes that applies the Bayes' theorem to train a probabilistic classifier [18].

Table 5 Model Training Dimensions

| Dimension | Description | Dimension Configurations |
|---|---|---|
| Datasets | Using different datasets to train the model | BeamNG.AI (AF 1,1.5,2), Driver.AI, and Combined Datasets |
| Features | Changing the features used for the model | Full Road Features, Road Segment Features |
| Training Set | Changing training set size by using different percentage split for training and test sets | 40% training set & 60% test set; 50% training set & 50% test set; 60% training set & 40% test set; 80% training set & 20% test set. |

We trained the ML models mentioned above using a training and test sets split strategy, for each of the configurations listed in Table 4, separately. We evaluated the performance of each ML model by computing the standard metrics of precision, recall, and accuracy [11, 14, 17, 19, 40, 49], computed as follows:

$$\text{Precision} = \frac{TP}{TP + FP}$$

$$\text{Recall} = \frac{TP}{TP + FP}$$

$$\text{Accuracy} = \frac{TP + TN}{TP + TN + FP + FN}$$

In the formula, we refer with TP the true positive cases (i.e., unsafe tests correctly identified), while with FP, the cases in which safe tests have been miss-classified as unsafe tests. Vice versa, in the formula, we refer with FP the true negative cases (i.e., safe tests correctly identified), while with FN, the cases in which unsafe tests have been miss-classified as safe tests.

Since unsafe scenarios are an exception {not the norm{ when generating random tests, the raw data we collected with AsFault are unbalanced toward safe cases. Therefore, we rebalanced the training data to avoid skewed distri-butions that would otherwise bias the ML models towards one specific class. Specifically, we adopted random oversampling, a rebalancing technique proven to be robust [44], to supplement the training data with multiple copies of some of the minority classes. To study how the training set size affects the ML mod-els' performance, we created balanced data sets of increasing size (Table 5). We generate the test sets necessary to evaluate the ML models by randomly sampling the data point not included in the training set. Notably, we did not rebalance the test set to preserve the underlying distribution classes in the data.

We also study the effects of different training strategies on each ML model' performance. To do so, instead of creating a balanced dataset for each configuration, we evaluated the ML models using standard K-fold cross-validation [51]. In particular, we set K = 10 and utilize all the available data in each configuration.

Table 6 Offline Experiment Dataset

| Dataset | Number of Safe Tests | Number of Unsafe Tests |
|---|---|---|
| Complete Set | 3095 | 2543 |
| Training Set | 2034 | 2034 |
| Test Pool (95/5) | 1061 | 55 |
| Test Pool (80/20) | 1061 | 265 |
| Test Pool (60/40) | 763 | 509 |
| Test Pool (30/70) | 218 | 509 |

### 3.2.2 Offline Experiments

To answer RQ2, we need to understand how frequently SDC-Scissor selects unsafe test cases and whether it devotes them more execution time than safe test cases. SDC-Scissor can use pre-trained models to classify safe and unsafe test cases. However, it can also retrain the ML models on the fly, as new data are collected from test executions. Therefore, we plan experiments to analyze how using pre-trained ML models for selecting (existing) test cases improves regression testing. Likewise, we plan experiments to assess the effectiveness of SDC-Scissor as a means to dynamically adding new test cases to automatically generated test suites (see Section 3.2.3). For those experiments, we consider the combinations of ML models and features that achieve the best results in the context of RQ1 (see Section ). Finally, we contextualize the results achieved by SDC-Scissor using a baseline approach that performs a random selection of test scenarios. Notably, random selection is considered one of the standard baselines for evaluating test selection strategies [55, 64].

Studying the effectiveness of SDC-Scissor offline requires test cases and executions; therefore, we used the data previously collected in the context of RQ1. Specifically, we consider the data generated while testing BeamNG.AI in the moderate configuration (AF 1.5). We decided so because this configuration provides a large number of test cases and executions (see Complete Set in Table 6). From the Complete Set, we created a Training Set, accounting for 80% of the data, and we used the remaining 20% of data for testing. We created a balanced Training Set, but we purposely created four unbalanced Test Pools with distribution of unsafe cases ranging from few (5% of the testing data) to many (70% of the testing data). Our conjecture is that using different Test Pool compositions allows assessing SDC-Scissor's performance in various settings.

We conducted the offline experiment in two experimental setups, referred to as FIX and REACH, and repeated the experiments in both setups thirty times to increase the confidence in the achieved results.

The FIX setup investigates the benefits of using SDC-Scissor when the resources allocated for testing are limited, i.e., the amount of test cases that can be executed in the simulation environment is fixed (e.g., S). The process we followed to experiment with the FIX setup is reported in Figure 2 alongside the baseline process. The baseline draws tests from the test pool at random and adds them to the test suite until the test suite reaches the target size S. FIX, instead, samples the tests from the test pool but adds them to the test suite only if the ML model predicts that they are unsafe; as before, the process ends when the test suite reaches the target size S. In this setup, more effective techniques select larger portions of unsafe tests; therefore, we evaluate the performance of SDC-Scissor using the ratio of unsafe to safe test cases in the final test suites.

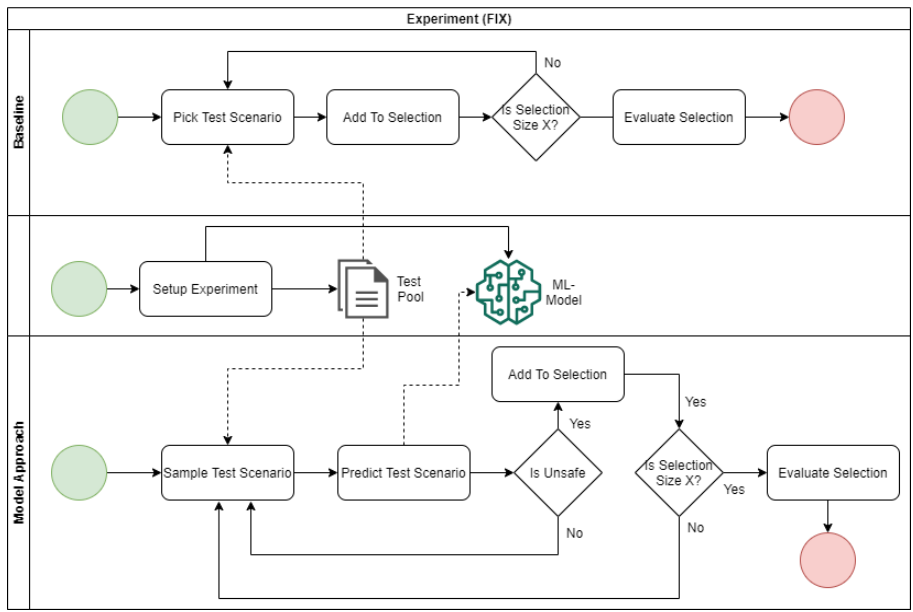

Fig. 2 FIX Experiment Overview.

The REACH setup investigates the ability of SDC-Scissor to reduce the time to identify at least N unsafe test scenarios. We conjecture that testing time should be spent on executing unsafe test cases, as those help developers expose problems of SDCs earlier. In our experiment, we set N = 10, since the time to identify that many unsafe test cases potentially requires the execution of many more (safe) test cases. The process we followed to experiment with the REACH setup is reported in Figure 3 alongside the baseline process. As before, the baseline randomly samples tests from the test pool and executes them until N unsafe tests have been identified. REACH, instead, follows a similar process but executes only those tests that are predicted to be unsafe by the ML models. In this setup, more effective techniques identify N unsafe tests sooner; therefore, we consider the number of true positives (TP),[5] true negatives (TN), false positives (FP), and false negatives (FN) predicted by the ML models. Having information about TP, TN, FP, and FN enables us to count how many tests were needed to reach the goal, how long it took to do so, and how much time was wasted in evaluating safe test cases.

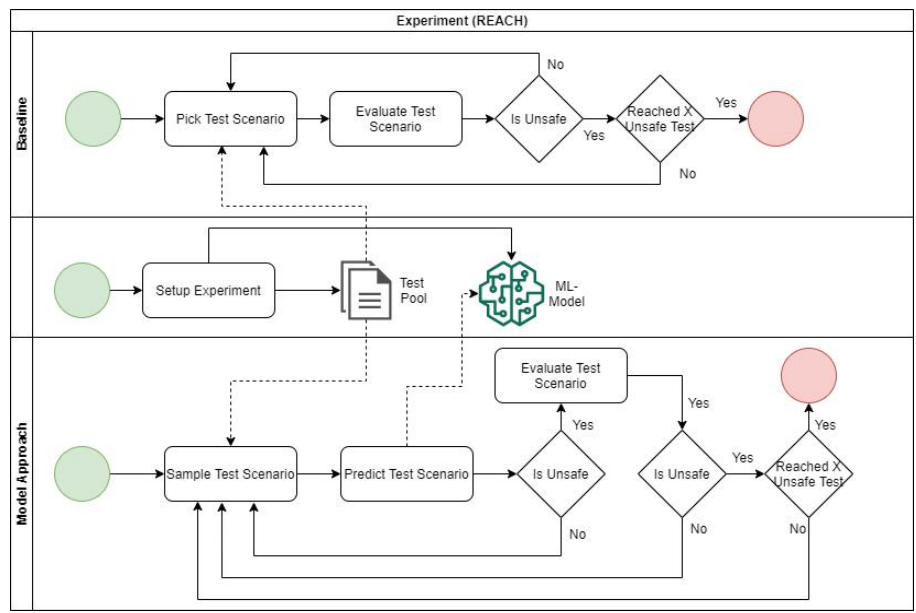

Fig. 3 REACH Experiment Overview.

### 3.2.3 Real-Time Experiments

We complement the previous Offline Experiments, which focus on applying SDC-Scissor to regression test case selection, with Real-Time Experiments, which study the application of SDC-Scissor to automated test generation.

We conducted the Real-Time Experiments according to the following pro-cedure: (i) AsFault to generates random test cases; (ii) for each newly gen-erated test case, SDC-Scissor classifies it as safe/unsafe; and, (iii) we filter out test cases classified as safe before generating the next test case, whereas we executed the test cases classified as unsafe. (iv) As test subject, we used BeamNG.AI in the moderate configuration (AF equal to 1.5) as this configura-tion is a compromise between overly conservative and overly aggressive driving styles.

A cost-effective test generator devotes more time to execute (likely) unsafe tests that can expose defects rather than executing safe test cases, which might not contribute any additional insight into the behavior of SDC under test. Cor-rectly identifying unsafe test cases, therefore, is paramount and depends on the quality of the ML model used as a classi er which, in turn, depends on the technique employed by the ML models and the data used to train them. Particularly relevant in this context is whether the ML model is prede ned and xed or allowed to be updated online as new data become available. The trade-o between these two con gurations is that ML models have little op-erational costs once trained but may miss relevant behaviors; on the contrary, dynamically retrained ML models can cope with missing training data but at the cost of additional time spent in retraining them. Therefore, we compare the following two approaches:

- Pre-trained Model in which we used the best performing model iden-ti ed during the Machine Learning-based Experiments (Section 3.2.1. We trained this model using the re-balanced dataset for the case of BeamNG.AI AF 1.5, as this is the con guration of the test subject used for this exper-iment.

- Adaptive Model in which we also used the best performing model iden-ti ed during the Machine Learning-based Experiments (Section 3.2.1 but trained with only 60 randomly generated test cases. After this initial train-ing, we retrain the ML model after executing the predicted unsafe test cases using the newly collected ground truth labels for those test cases. Figure 4 illustrates this process. Notably, since the ML model may be inaccurate, this process collects both positive and negative labels.

---

[5] True positives are tests predicted as unsafe and verified to be so; conversely, true nega-tives are tests predicted and verified to be safe.

As before, we contextualize the results achieved by SDC-Scissor using a baseline approach that implements plain vanilla random generation, i.e., it does not lter the test cases.

We ran each con guration on a dedicated machine equipped with an Intel Core i5-6600K (3.5 GHz), 16 GB RAM, and an NVIDIA GeForce GTX 1070 GPU, and set the test generation time budget to six hours.

During each execution of the experiment, we stored all the tests generated by AsFault so we can execute the test cases ltered out by SDC-Scissor post-mortem to calculate metrics such as accuracy, precision, and recall.

Table 7 provides an overview of the metrics used for the evaluation of SDC-Scissor across the various con gurations. Those metrics include the count of unsafe tests found during each experiment (true positives), true negatives, false positives, and false negatives. Additionally, we consider how SDC-Scissor allocated the time budget to run safe and unsafe test cases, generate test cases, and rebuild the ML models.

Finding .

## 4 Results

This section reports, for each research question, the obtained results and the main ndings.

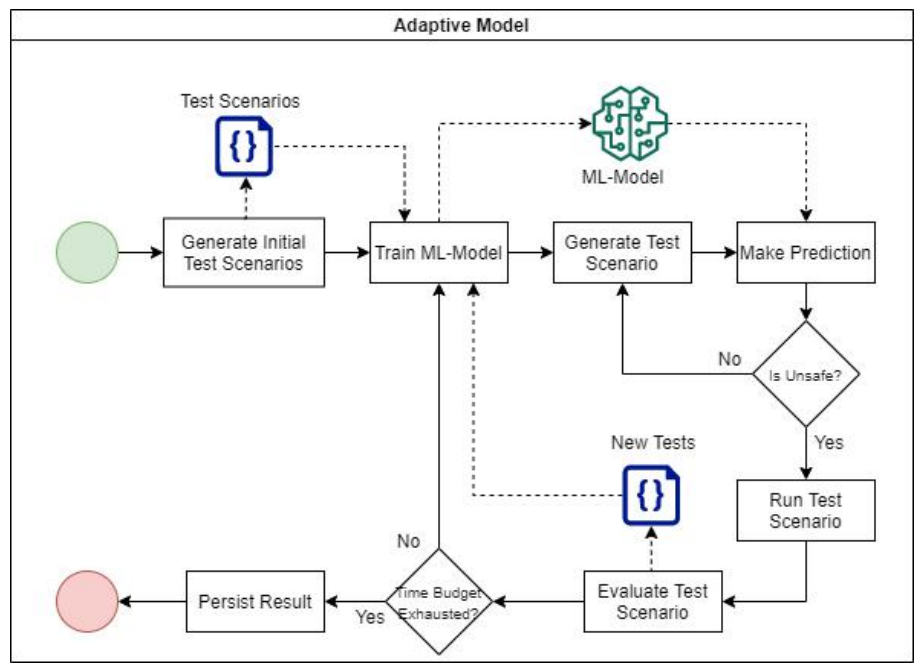

Fig. 4 Overview of the Adaptive Model configuration for the Real-Time Experiments.

Table 7 Evaluation metrics for the Real-Time Experiments.

| Metric | Description | Range |
|---|---|---|
| Number of Unsafe Test Execution | The number of unsafe tests the approach simulated during the experiment | 0-N |
| Number of Safe Tests Execution | The number of safe tests the approach simulated during the experiment | 0-N |
| Time Allocation | How much time relative to the total time was spent with an action | 0-1 |
| True Positives/Negatives | Number of correct prediction for categories safe and unsafe | 0-Number of Predictions |
| False Positives/Negatives | Number of incorrect prediction for categories safe and unsafe | 0-Number of Predictions |

## 4.1 Machine Learning-based Experiments

In this section, we discuss the results of RQ1. Specifically, we first describe the results achieved using Full Road (Section 4.1.1) and Road Segment features (Section 4.1.2) to build the ML models; next, we describe the effects of using various training and test configurations (Section 3).

### 4.1.1 Full Road Features

We evaluated the ML models trained using Full Road features with four different splits of training and test data (see Table 5). However, for the sake of readability, we only report the ML models' performance metrics (i.e., accuracy, precision, recall, and F1 score) of the best performing configuration (i.e., 80% training and 20% for testing) in Table 8. The full results can be found in our replication package [41]. In the table, we report an aggregate value of accuracy (column Acc.), but we present precision, recall, and $F_1$ score separately for unsafe and safe labels. We do so because the test set follows the original distribution of safe and unsafe test cases; hence, it is unbalanced and highly biased towards safe cases. Consequently, reporting an aggregated value for those metrics may not be representative as the strong presence of safe cases would dominate the results.

Table 8 Performance of the ML models trained using full road features. The results refer to the split 80/20 between training and test data. The best results are shown in bold face.

| Model | Acc. | Unsafe | | | Safe | | |
|---|---|---|---|---|---|---|---|
| | | Prec. | Recall | $F_1$ | Prec. | Recall | $F_1$ |
| BeamNG AF 1 | | | | | | | |
| J48 | 61.2% | 37.6% | 69.8% | 48.9% | 84.2% | 58.0% | 68.7% |
| Naïve Bayes | 55.7% | 36.7% | 92.1% | 52.5% | 93.7% | 42.5% | 58.5% |
| Logistic | 66.2% | 43.3% | 87.3% | 57.9% | 92.7% | 58.6% | 71.8% |
| Random Forest | 63.7% | 40.7% | 79.4% | 53.8% | 88.6% | 58.0% | 70.1% |
| BeamNG AF 1.5 | | | | | | | |
| J48 | 65.6% | 69.2% | 67.4% | 68.2% | 61.5% | 63.5% | 62.5% |
| Naïve Bayes | 66.7% | 79.3% | 53.2% | 63.6% | 59.3% | 83.1% | 69.2% |
| Logistic | 70.9% | 78.1% | 65.3% | 71.1% | 64.8% | 77.8% | 70.7% |
| Random Forest | 68.5% | 75.8% | 62.7% | 68.6% | 62.5% | 75.6% | 68.4% |
| BeamNG AF 2 | | | | | | | |
| J48 | 90.8% | 98.7% | 91.5% | 95.0% | 28.2% | 73.3% | 40.7% |
| Naïve Bayes | 93.4% | 98.7% | 94.3% | 96.4% | 36.7% | 73.3% | 48.9% |
| Logistic | 83.2% | 99.6% | 82.8% | 90.4% | 19.7% | 93.3% | 32.6% |
| Random Forest | 92.8% | 99.7% | 92.7% | 96.1% | 36.8% | 93.3% | 52.8% |
| Driver.AI | | | | | | | |
| J48 | 44.1% | 19.5% | 64.1% | 29.9% | 82.9% | 39.6% | 53.6% |
| Naïve Bayes | 38.8% | 20.3% | 78.5% | 32.3% | 85.8% | 29.8% | 44.2% |
| Logistic | 56.3% | 22.7% | 56.5% | 32.4% | 85.0% | 56.3% | 67.7% |
| Random Forest | 57.2% | 22.3% | 52.6% | 31.3% | 84.4% | 58.2% | 68.9% |

Regarding the BeamNG.AI datasets, we can observe that the ML models' accuracy improved for increasing AF levels. For instance, with AF2 SDC-Scissor reached a precision of 99.7% for unsafe predicted tests. The dataset composition seems to be the key factor explaining this result since setting the aggression factor to higher values resulted in significantly more unsafe cases. Conversely, a small number of safe cases improved accuracy and precision for unsafe cases counterbalanced by a decrease in the precision of safe predictions.

Finally, we can observe a similarity between the ML models' $F_1$ score for safe and unsafe classes for the BeamNG.AI AF 1.5 case. This result can be explained by looking at how evenly distributed the safe and unsafe classes are, which illustrates the importance of having unbiased datasets for training and testing the models.

We can also observe that the ML models achieved lower accuracy for Driv-ing.AI (49.1%) than BeamNG.AI AF 1.5 (accuracy 67.9%). This result can be explained by looking at how unbalanced the Driver.AI dataset is. Since Driver.AI drives carefully, its dataset comprises mainly safe scenarios, and the predictions of the ML models tested on it are biased towards safe predictions. Comparing the $F_1$ score achieved by the ML models against the Driver.AI and BeamNG.AI AF 1.5 datasets shows this problem more evidently: the ML models performed comparably well for safe and unsafe classes against the BeamNG.AI dataset, whereas they performed well only for the safe test class in the case of Driver.AI. This result supports the observation that the more the SDC under test drives safely, the harder it becomes to predict unsafe test cases.

Finding 1. The accuracy of SDC-Scissor is in uenced by the driving agents, their driving style, and the diversity of datasets. For example, for more aggressive driving agents, the accuracy achieved by the ML models was higher than for cautious driving agents. Hence, predicting unsafe test cases is harder for cautious drivers than reckless ones. Consequently, improving testing of SDCs is more challenging for less aggressive driving agents.

We studied the ability of the ML models to transfer knowledge from a driving agent to another one by using the BeamNG AF 1.5 dataset to train the ML models but using the Driver.AI test set, generated from the same set of virtual roads, to evaluate them, and vice versa. As it is possible to observe in Table 9 the knowledge from a driving agent is not transferable to another one. Table 9 shows that the ML models trained on Driver.AI and evaluated on BeamNG performed signi cantly worse than the same models trained on BeamNG exclusively (from 67.9% to 41% on average). However, when training the ML models on the BeamNG.AI dataset and evaluating them using the Driver.AI datasets, the ML models performed only slightly worse (between 49.1% and 47.8% on average). Interestingly, when using both datasets together, the results show a compromised solution between the accuracy achieved when training on the di erent AI engines separately: BeamNG 67.9%, Driver.AI 49.1%, Combined datasets 55.5%.

Finding 2. Our results show that the knowledge is not transferable from one driving agent to another, i.e., SDC-Scissor performed worse when training ML models on data from a speci c driving agent and testing them on data from a di erent one. Despite this, ML models trained on

Table 9 ML Models' accuracy on mixed datasets.

| Model | Training Acc. | Test Acc. |
|---|---|---|
| BeamNG/Driver.AI (Training/Test) | | |
| J48 | 87% | 46% |
| Naive Bayes | 67% | 56% |
| Logistic | 72% | 45% |
| Random Forest | 100% | 44% |
| Driver.AI/BeamNG (Training/Test) | | |
| J48 | 84% | 44% |
| Naive Bayes | 66% | 35% |
| Logistic | 81% | 45% |
| Random Forest | 100% | 43% |
| Driver.AI & BeamNG Combined | | |
| J48 | 71% | 53% |
| Naive Bayes | 61% | 49% |
| Logistic | 64% | 60% |
| Random Forest | 87% | 56% |

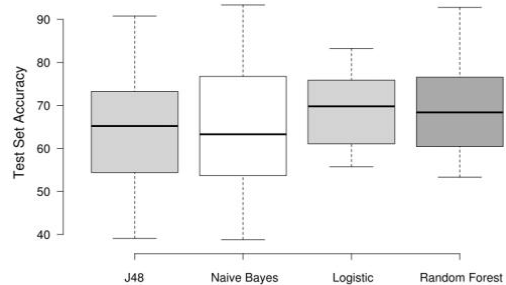

Fig. 5 Comparing test accuracy for di erent machine learning models for all datasets. J48 (median=65.2%), Naive Bayes (median=63.3%), Logistic (median=69.8%), Random Forest(median=68.4%)

the BeamNG data performed only slightly worse when evaluated on the Driver.AI data.

We are interested in nding the most suitable ML model to be used by SDC-Scissor; therefore, we consider the accuracy achieved by the various ML models across all datasets. As can be seen from Figure 5, the various ML models achieved comparable accuracy. The Logistic model has the highest median value (69.8%) but did not perform drastically better than the worst performing model, i.e., the Naive Bayes model (63.3%). However, the Logistic model's accuracy seem to be more stable, as the variability associated to its results is smaller than the other models.

Table 10 Performance of the ML models trained using road segment features. The best results are shown in bold face.

| Model | Acc. | Unsafe | | | Safe | | |
|---|---|---|---|---|---|---|---|
| | | Prec. | Recall | F₁ | Prec. | Recall | F₁ |
| BeamNG AF 1.5 | | | | | | | |
| J48 | 82.4% | 3.5% | 88.8% | 6.7% | 99.9% | 82.3% | 90.3% |
| Naive Bayes | 53.2% | 1.2% | 82.0% | 2.4% | 99.8% | 53.0% | 69.2% |
| Logistic | 72.1% | 1.9% | 77.0% | 3.8% | 99.8% | 72.1% | 83.7% |
| Random Forest | 84.9% | 4.1% | 90.0% | 7.9% | 99.9% | 84.8% | 91.8% |
| Driver.AI | | | | | | | |
| J48 | 77.8% | 0.8% | 79.4% | 1.6% | 99.9% | 77.8% | 87.5% |
| Naive Bayes | 66.1% | 0.4% | 58.4% | 0.8% | 99.9% | 66.1% | 79.5% |
| Logistic | 62.5% | 0.5% | 79.9% | 1.0% | 99.9% | 62.5% | 76.9% |
| Random Forest | 79.1% | 0.9% | 81.3% | 1.8% | 99.9% | 79.1% | 88.3% |

Finding 3. No machine learning model outperformed the others in terms of accuracy. However, among them, the Logistic model provided the more stable results.

### 4.1.2 Road Segment Features

We investigated the use of local features describing road segments for training ML models that can accurately predict whether test cases are safe or unsafe. However, training the ML models using road segment features showed opposite results when predicting safe and unsafe test cases. As Table 10 shows, the ML models achieved very high precision while predicting safe test cases but were imprecise while predicting unsafe test cases. One possible explanation of this behavior is that virtual roads mainly consisted of "safe" road segments, i.e., road segments that belong to roads in which no OBE was observed. Hence, the number of safe road segments was significantly higher than the number of unsafe ones. As a result, the ML models were biased and consistently favored safe predictions in all experiments.

Finding 4. Although using road segment features to train ML models achieved better accuracy (85%) than using full road features, the ML models trained using road segment features achieved very low precision for the critical, unsafe class. The high accuracy of ML models trained using road segment features is an artifact of the strong bias towards safe cases of the data and has no practical benefits in SDC testing optimization where the goal is to invest more effort in executing unsafe test cases.

### 4.1.3 Analysis of Relevant Features.

In our study, we considered two sets of features, full road, and road segment features. Although the ML models trained using these feature sets can effectively classify the test cases as safe or unsafe, it is crucial to know the contribution of each of these features. For instance, more profound knowledge of the features may help to de ne better-suited feature sets. Hence, we analyzed in detail the full road features using the BeamNG AF 1.5 dataset.

Table 11 Feature Selection Rankings according to A) Information Gain Analysis, B) Correlation Analysis

| | A | | | B | |
|---|---|---|---|---|---|
| Rank | Feature | Inf. Gain | Rank | Feature | Correlation |
| 1 | min pivot o | 0.140 | 1 | min pivot o | 0.342 |
| 2 | mean pivot o | 0.087 | 2 | total angle | 0.332 |
| 3 | total angle | 0.085 | 3 | num l turns | 0.330 |
| 4 | num l turns | 0.084 | 4 | mean pivot o | 0.326 |
| 5 | num r turns | 0.077 | 5 | num r turns | 0.316 |
| 6 | std pivot o | 0.067 | 6 | std pivot o | 0.270 |
| 7 | median pivot o | 0.050 | 7 | median pivot o | 0.257 |
| 8 | length | 0.039 | 8 | length | 0.222 |
| 9 | num straights | 0.013 | 9 | num straights | 0.138 |
| 10 | std angle | 0.011 | 10 | max angle | 0.109 |
| 11 | max angle | 0.011 | 11 | min angle | 0.104 |
| 12 | min angle | 0.010 | 12 | max pivot o | 0.063 |
| 13 | max pivot o | 0.003 | 13 | direct distance | 0.053 |
| 14 | direct distance | 0.003 | 14 | std angle | 0.048 |
| 15 | median angle | 0.002 | 15 | median angle | 0.025 |
| 16 | mean angle | 0.000 | 16 | mean angle | 0.017 |

Table 11 reports the results of using two popular feature evaluation meth-ods: information gain and correlation. We order the features based on their evaluation scores and set a threshold (0.01 for information gain and 0.1 for correlation) for each evaluation method to select only the features with the highest contribution. It can be seen from Table 11-A and Table 11-B that the ordering and the relative score of the features are similar in most of the top cases among the two methods. Speci cally, the top eight features are precisely the same in both methods, with a slight change in the order between ranks 2 to 4. Additionally, we note that the remaining features above the thresholds di er in just one feature, i.e., "std angle" which ranked in correlation score lower than the information gain (rank 14 vs. 10).

Overall, we observe that almost all full road features contributed to distin-guish safe versus unsafe test cases. Also, among the statistical features that we reported in Table 2, features concerning the pivot radius tend to be more critical and relevant for the distinction of the classes. The minimum and av-erage radius of the pivots are among the most contributing features, while the statistics concerning the turn angles start appearing only from rank 10.

Finding 5. The de ned features contribute di erently to characterizing the safe and unsafe scenarios. The statistics concerning the pivot radius (min, mean, std, median), the sum of the turn angles, the number of left and right turns, and the total length of the road are among the most
important features, which are all belonging to the set of full road features.

## 4.2 Offline Experiments

In this section, we discuss the results of RQ2. Specically, we report the
results of the FIX and REACH experiments (detailed in Section 3).
Additionally, we report the results of the comparison between various ML
models against the baseline approach (described in Section 3) by
considering dierent test pool compositions.

### 4.2.1 FIX Experiment results

The goal of this experiment is to optimize the usage of the available resource in
terms of testing execution time and e ectiveness. Figure 6 compares the ratio of
unsafe tests selected for execution using dierent ML models against the
baseline approach (random selection) across dierent test pool composi-tions.
As can be observed from the gure, the Logistic model outperformed the
baseline in all dierent test pool compositions (described in Section 3). Figure 7
illustrates that with fewer unsafe test cases in the pool, we observe
improvements in the number of selected unsafe tests using ML models over the
baseline. In the pool with the least unsafe tests, the Logistic model nds 133%
more unsafe test tests compared to the baseline approach. In the more bal-
anced testing pool, Logistic nds 50% more unsafe tests, while with the pool with
more unsafe than safe tests, it identies 30% more unsafe tests. The Lo-gistic
model performs slightly better than the other models in all compositions except
one (0.3/0.7) where Random Forest performed the best.

   The confusion matrices in Table 12 further illustrate the concrete results
in terms of e ectiveness with the various pool compositions. In the pool with
only 0.05 unsafe tests (Table 12-a), the Logistic model achieved 10 false
negatives and 260 true negatives; this means that the model avoided the
execution of 549 safe test tests (considering that safe test cases in average
take around 24 seconds in average to be executed), thus potentially
reducing cost by more than 200 minutes in total on the less critical scenario.
However, the false-positive number is still high, with a cumulative 263 false-
positive identied. As can be observed in Table 12)-b, for the Test Pool
0.7/0.3 the Logistic model achieved over 260 true positives and only 37
false positives. We observe that the precision correlates with the dataset
composition, indeed, for datasets having more unsafe tests, the precision
for unsafe tests is higher. For datasets having less unsafe tests, we obtain
the opposite e ect in the results. Figure 6 shows that the ML model
performance and the baseline depend on the test compositions. The
baseline and ML models perform better in test pools with more unsafe test
tests. Thus, according to our results, designing an appropriate test pool
composition is of critical importance to achieving accurate prediction results.

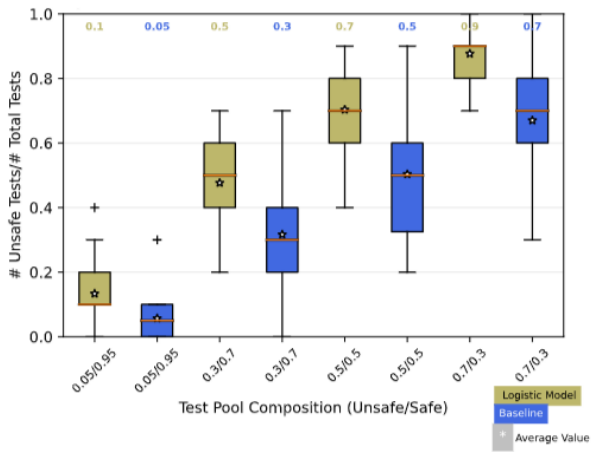

Fig. 6 Comparison Logistic Model and Baseline across different Test Pool Compositions.

Fig. 7 Number of executed unsafe scenarios during the experiments on a)Test Pool(0.05/0.95) b) Test Pool(0.3/0.7) c)Test Pool(0.7/0.3)

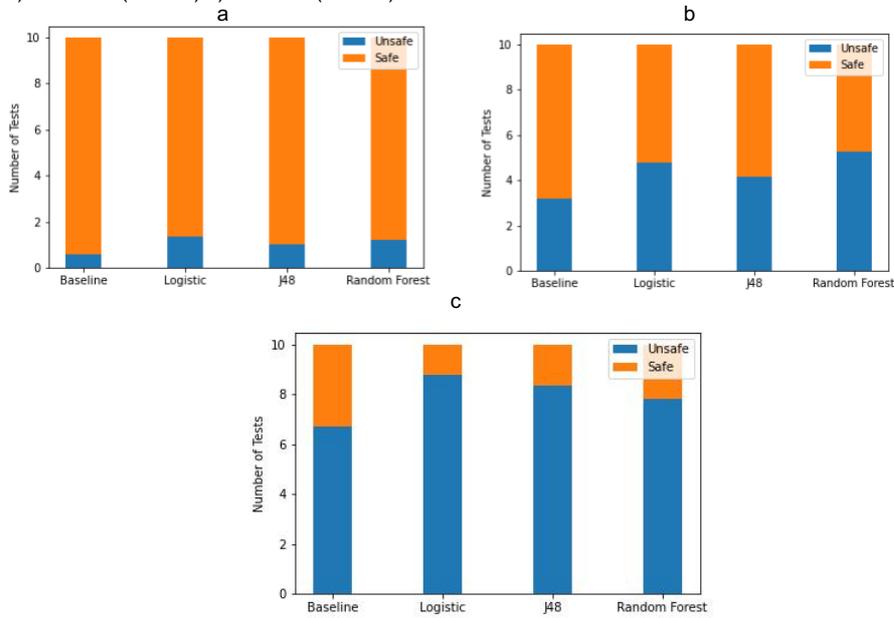

Finding 6. SDC-Scissor outperforms the baseline approach in selecting unsafe tests across all test pool compositions, critical for more effective testing practices. In the test pool composition 0.3/0.7 (safe to unsafe) SDC-Scissor was able to find 30% more unsafe tests and in the test pool

Table 12 Confusion Matrix for Logistic Model,Cumulative over 30 rounds for a) Test Pool (0.05/0.95), b) Test Pool (0.7/0.3)

a

| | | Predicted class | |
|---|---|---|---|
| | | Unsafe | Safe |
| Actual Class | Unsafe | 40 | 10 |
| | Safe | 260 | 549 |

b

| | | Predicted class | |
|---|---|---|---|
| | | Unsafe | Safe |
| Actual Class | Unsafe | 263 | 48 |
| | Safe | 37 | 81 |

composition 0.95/0.05 (safe to unsafe) composition it found 133% more unsafe tests.

## 4.2.2 REACH Experiment

The goal of this experiment is to investigate whether the usage of ML models allows reducing the total test execution time. By reducing the total test execu-tion costs, a testing pipeline would be able to spend more testing time on more safety-critical test cases. The task in this experiment was to identify, as early as possible, 10 unsafe tests while minimizing the number of total executed test cases. To perform the various comparisons, for each experimented strategy, we collected the information about the number of test cases required to reach 10 unsafe cases as well as the cumulative cost (i.e., the execution time) to run all the test cases (i.e., till the final unsafe scenario was identified). Further, we collected information concerning the execution time for both safe and un-safe test cases. The conjecture behind this analysis is that the testing cost concerning safe cases should be limited as much as possible, whereas the test cost dedicated on unsafe cases is bene cial to identify aws of SDC in virtual environments.

Figure 8 and Figure 9 provide an overview of the performance of the base-line compared to the Logistic model (the best performing model in previous experiments) across di erent test pool compositions. Table 13 summarize the results for the REACH experiment. We observed that the Logistic model per-formed better across all test pool compositions. The test costs strictly depend on the required numbered of tests to be executed before identifying the mini-mal set of 10 unsafe tests. Although the di erence in the number of required tests tends to be higher in the pool with less unsafe tests (in the 0.05/0.95 pool between 171 to 98.5 tests, in the 0.7/0.3 between 14 to 11 tests), SDC-Scissor allows reducing test execution time dedicated to less critical tests when the test pool presents more unsafe tests. Figures 10 show that in the smaller unsafe pool it is higher the test execution time dedicated to less critical tests. The test execution time to these less critical tests is 85% higher in the baseline than in the Logistic model. In the larger pool, the Logistic model reduces the unnecessary execution time by 170%.

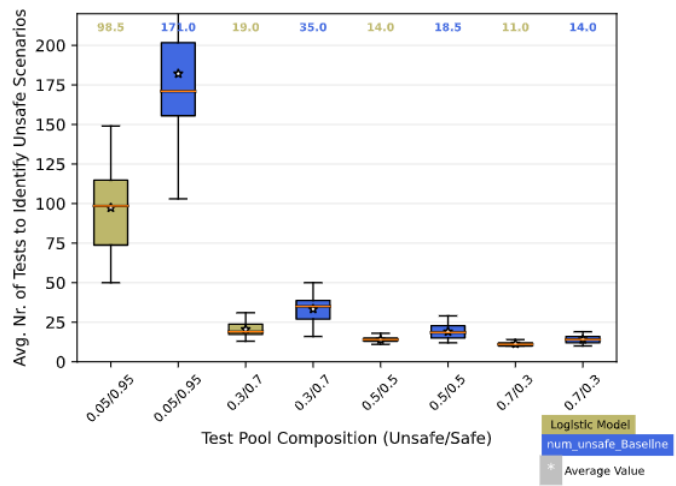

Fig. 8 Comparing the Logistic model with the baseline across the di erent test pools.

Table 13 Results of the REACH experiments comparing the Logistic model and the base-line. Execution time is reported seconds and the values are averaged across the experiment repetitions.

| Model/Pool | Tests # | Execution Time | |
| --- | --- | --- | --- |
| | | Safe | Unsafe |
| Smart Selector | | | |
| Test Pool (0.05/0.95) | 98.5 | 4664 | 375 |
| Test Pool (0.3/0.7) | 19 | 475 | 376 |
| Test Pool (0.5/0.5) | 14 | 214 | 389 |
| Test Pool (0.7/0.3) | 11 | 54 | 379 |
| Baseline | | | |
| Test Pool (0.05/0.95) | 171 | 8079 | 382 |
| Test Pool (0.3/0.7) | 35 | 1243 | 383 |
| Test Pool (0.5/0.5) | 18.5 | 439 | 391 |
| Test Pool (0.7/0.3) | 14 | 193 | 387 |

Finding 7. We investigate whether SDC-Scissor can reduce the number of executed tests required to nd at least N unsafe tests. Our results show that SDC-Scissor outperformed the baseline across all test pools, with the Logistic model reducing the unnecessary execution time dedicated to safe tests by 170%. SDC-Scissor performed better compared to the baseline when test pools are characterized by fewer unsafe tests.

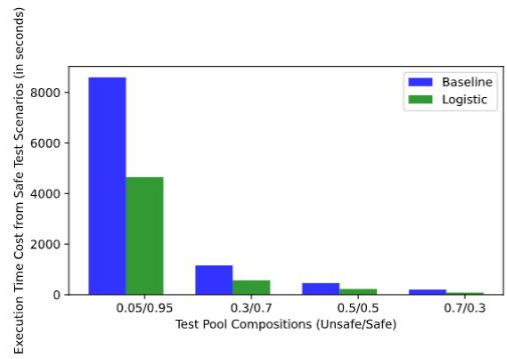

Fig. 9 Time spent for the execution of safe tests, Logistics vs. Baseline across different test pools

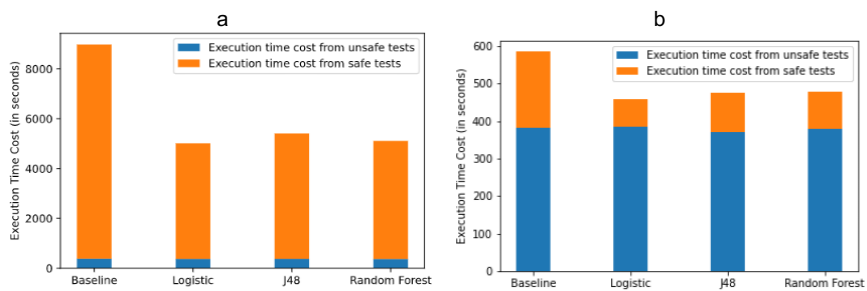

Fig. 10 Time spent on executing each safe and unsafe test cases for different models in a)test pool (0.7/0.3) b) test pool (0.05/0.95)

## 4.3 Real-time Experiment (RQ2)

In this section, we discuss the results of the real-time experiments, where we compare the results of a pre-trained model and a real-time model with the baseline approach, as described in Section 3.

Baseline vs. Pre-trained and Adaptive Models. Figure 11 gives an overview of the results achieved by the experimented models. We observe that the baseline executed the higher number of test cases (472). The pre-trained model runs more test cases (405) than the real-time approach (378). Figure 11 summarizes our main observations, as elaborated in the next paragraphs.

The pre-trained and real-time models apply a machine learning-based test selection, which leads to numerous rejected (i.e., non-executed) test cases: real-time and pre-trained experienced 588 and 309 rejected tests respectively. The baseline uses 98% of the time to test cases; only 2% are dedicated to generating test cases. The pre-trained and real-time approaches use more time for test generation (6% pre-trained, 11% real-time approach). In addition to the longer test generation process, these two approaches allocate time for predictions and evaluation of tests (pre-trained 4%, real-time 5%), which the baseline does not need to perform. Compared to the pre-trained approach, the real-time approach continuously trains the machine learning model with new tests.

Table 14 Comparison between pre-trained and real-time models.

| Model | Acc. | Unsafe | | Safe | |
|---|---|---|---|---|---|
| | | Prec. | Recall | Prec. | Recall |
| Pre-trained Model | 72.1% | 65.2% | 82% | 81.2% | 64% |
| real-time Model | 69% | 67.7% | 59.3% | 69.9% | 77% |

Interestingly, although the baseline execute more test cases, both pre-trained and real-time approaches found more unsafe test cases (baseline 195, pre-trained 265, real-time 256). The pre-trained model was able to nd 35% more unsafe test cases, executing only 49% safe tests. In Figure 11, we can observe that the baseline only spends 34% of the time running unsafe tests, while 64% of the test time was spent on executing safe test tests. In contrast, our proposed approaches dedicated more than 50% of the time on unsafe tests, which is positive, since, in a testing environment, the goal is to nd more er-rors in less time (in our case, it corresponds to expose more weakness in SC critical tests).

Finding 8. Our results show that even though the baseline approach executes more test cases, both the real-time and the pre-trained (i.e., o ine) models integrated into SDC-Scissor are able to nd more unsafe tests than the baseline. The time investment of predicting the outcome of test cases and generating more tests is bene cial for testing purposes. The pre-trained model was able to nd 35% more unsafe tests than the baseline, with the baseline only dedicating 34% of the time budget on assessing unsafe tests. The o ine model spends 52% running unsafe, and only 38% safe test cases.

Adaptive vs. Pre-trained Model. Figure 11 shows that the testing time alloca-tion for the pre-trained and real-time models is similar, but the real-time model spends more time for test generation (11%) than the pre-trained one (6%). The pre-trained model is based on the previously generated dataset with 5,643 test cases (as described in Section 3), whereas the real-time model started with generating an initial dataset of 60 test cases as described in Section 3. Table 14 shows that the pre-trained model achieved a higher accuracy (72.1%) than the real-time model (69%). The lower accuracy explains the higher number of test cases generated by the pre-trained model (tests generated; real-time 962, pre-trained 714). Although the pre-trained model has higher accuracy in general and higher unsafe recall, it only found 3.13% more unsafe tests than the real-time model.

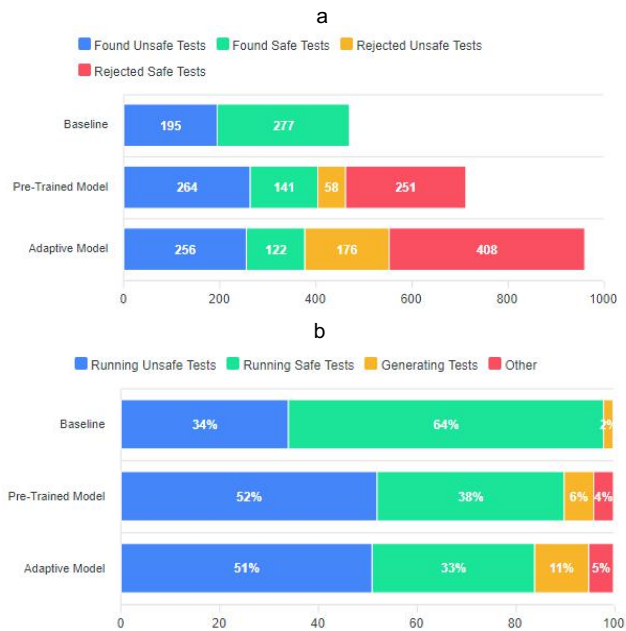

Fig. 11 Comparison of the metrics for different real-time approaches in a 6-hour run a) generated test cases distribution. b) spent time distribution across different tasks.

> Finding 9. The offline model achieved an accuracy of 72.1%, which is higher than the real-time model (69%). A real-time approach can achieve similar results compared to an offline model, with the real-time model finding only 3.13% fewer unsafe tests than the offline model. In achieving such results, the real-time model only used an initial set of 60 test cases, whereas the offline model leveraged 5,643 tests.

## 5 Related Work

In this section, we briefly discuss relevant research concerning the following topics: (i) academic and industry studies concerning DevOps limitation and practices in the context of general CPSs; (ii) studies proposing, using, or eval-uating simulation environments for testing SDCs.

### 5.1 Limitations of DevOps for CPSs

Contemporary DevOps pipelines allow improving the communication between Development(Dev) and Operations(Ops) during the software development pro-cess, enabling continuous improvements to existing and new products [35]. Sev-eral researchers and practitioners advocate DevOps as a promising approach for CPSs development [23, 61]. However, both in traditional [23, 66] and CPS application domains, the state-of-the-art of DevOps is still forming [61], and emerging practices need validation in the wild.

A recent survey by Torngren and Sellgren [61] discusses how CPSs' engineering deals with the inner complexity of CPSs' design and the challenges that arise from the environments in which CPSs operate. According to them, while (semi-automated) integration happens through software, there are several distinguishing characteristics between software and physical systems that make co-designing hardware and software hard. Those characteristics include entirely different approaches, techniques, abstractions, platforms, faults & failure modes, and development practices. They conclude that to cope with the foreseen demand of CPSs at scale and in multiple domains, CPS development and testing need rapid prototyping, code/test generation, and various testing phases [57] encapsulating X-in-the Loop ('XiL) activities. In XiL, the 'X' indicates the target of development and testing and typically refers to model (MiL), software (SiL), and hardware (HiL). A typical CPS development pipeline needs to efficiently and effectively integrate various XiL activities to support development and evolution [57, 61].

In this paper, we investigated ways to improve XiL activities, by focusing on proposing simulation-based test scenario generation and selection approaches for SDCs, equipped with machine-learning-based strategies for enabling the selection of most relevant (or critical) test cases (or scenarios). An optimal test case selection needs to target the identification of relevant (i.e., potentially unsafe) test cases, while optimizing simulated testing pipeline, to ensure higher safety for the final products.

## 5.2 Simulation-based Testing of SDCs

Given the danger and ineffectiveness of physical testing [39], researchers and practitioners devised simulation-based testing approaches for SDCs. This provides the opportunity to conveniently test the control software of a SDC in a (realistic enough) simulated environment, and in a diverse set of (automatically generated) test cases. Additionally, using simulation, the systems under test can face tests that might be otherwise too expensive, too hard, too risky, or impossible to recreate in real life [25].

Consequently, simulation is becoming one of the cornerstones in developing and validating SDCs, as it is heavily utilized in various XiL activities and across the entire development life-cycle. Simulation is currently used to support the initial inception and requirement analysis (MiL), Hardware/Software Co-design (MiL), design and testing of software components (SiL), training and validation of Machine Learning components (SiL), and testing and validation of the deployed system (HiL).

Abdessalem et al. [2] proposed an approach for test scenario generation for SDCs based on a combination of evolutionary search algorithms and decision tree classification models. Their goal is to leverage classification models to guide the search-based generation of tests faster towards critical test cases.

Also, search algorithms refine classification models so that the models can accurately characterize critical regions (i.e., the regions of a test input space that are likely to contain the most critical test cases). They evaluate their approach by generating test cases for an automated emergency braking system, and use PreScan [38], a commercial SDC simulator for the test execution.

AV-FUZZER was proposed by Li et al. [43], which consists of a testing framework to find the safety violations of an autonomous vehicle in the pres-ence of an evolving traffic environment. They leverage domain knowledge of vehicle dynamics and genetic algorithms to minimize the safety potential of an autonomous vehicle over its projected trajectory and design a local fuzzer that increases the exploitation of local optima in the areas where highly likely safety-hazardous situations are observed. For evaluating their proposed framework, They use an Unreal Engine based real-time simulation platform, LGSVL [26], that is capable of simulating complex urban and freeway driving scenarios using a library of urban layouts, buildings, pedestrians, and vehicles.

Search-based testing struggles to generate test scenarios with peculiar fea-tures, e.g., a rear-end car crash, that developers might need to validate or debug their implementations, and manually creating such test scenarios is time-consuming and cumbersome. Hence, Gambi et al devised AC3R [30], an approach to derive scenario-based tests from simulations of real car crashes. AC3R leverages natural language processing and a custom ontology to ex-tract information from police reports that describe car crashes and uses basic kinematics to plan the intercepting vehicles' trajectories. It uses BeamNG to automatically simulate the whole environment that re-enacts the car crashes.

As mentioned before in Section 2, we leveraged AsFault [31] to generate test case and BeamnNG [13] to run them in simulation in our study. Com-plementary to such previous studies, we specifically focused on designing and integrating an approach in the SDC testing pipeline, to recognize and exclude safe scenarios without executing them based on machine learning models. This research direction is relevant to save valuable testing and processing resources, and allocate computing power and time to execute more critical (i.e., unsafe) and potentially risky test cases.

## 6 Threats to Validity

Threats to internal validity may concern, as for previous work [32], the cause-effect relationships between the technologies used to generate the scenarios and their elements and the corresponding results, which strictly depends on the realism of our scenarios. Indeed, since we used AsFault, we did not recre-ate all the elements that can be found in real roads (e.g., weather condition, weather conditions, etc.). However, to increase our internal validity, we used both BeamNG.AI and Driver.AI as test subjects. They both leverage a good knowledge of the roads, which means that they do not suffer from limitations of vision-based lane-keeping systems. For future work, we plan to leverage the new BeamNG features, which allow to experiment with test cases composed by traffic lights as well as other cars and static objects.

Finally, threats to external validity concern the generalization of our findings. Although the (i) number of experimented test cases in our study is relatively larger [32]; and (ii) we experimented with different AI-engines (i.e., BeamNG.AI and Driver.AI) compared to previous studies; we cannot claim that our results can be generalized to the universe of general open-source CPS simulation environments in other domains. Therefore, further replications are desirable, so are further studies considering more data as well as other CPS domains. To further minimize potential external validity, in conducting our experimental evaluation, we followed the guidelines by Arcuri et al. [6] that suggests to compare results with randomized test generation algorithms (our baseline approach in RQ2) and repeated the experiments several times.

## 7 Conclusions and Future Work

Regression testing for SDCs is particularly challenging due to the cost of run-ning many test driving scenarios. To improve the cost-effectiveness of regres-sion testing, we introduced a test case selection approach, called SDC-Scissor, that relies on a set of SDC road features extracted from driving scenarios prior to running the tests in the context of BeamNG SDC simulation environment. Then, SDC-Scissor uses ML approaches to select the test cases having a higher likelihood to experience unsafe situations.

We empirically investigated the performance of SDC-Scissor and compared it with baseline approaches. Our assessment of SDC-Scissor shows that SDC-Scissor successfully selects test cases independently from the AI engine used or different risk levels (i.e., different driving styles), with the Logistic model providing the more stable results. Interestingly, our results also show that the knowledge is not transferable from one AI engine to another one, i.e., SDC-Scissor performed worse when training ML models on data from a specific AI engine and testing on data from a different AI engine. Moreover, among the defined features to train the ML models, the one that contribute the most in the accuracy of SDC-Scissor are the concerning the set of full road features.

Our findings also suggest that SDC-Scissor can reduce the number of exe-cuted tests required to find at least 10 unsafe tests. Specifically, SDC-Scissor outperformed the baseline across all test pools, with the Logistic model re-ducing the unnecessary execution time dedicated to safe tests by 170%. In terms of running time, we observed that is able to select test scenarios in a cost-effective manner compared to a random baseline approach.

As future work, we plan to replicate our study on further SDC datasets, AI engines, and SDC features. Moreover, we plan to perform new empirical studies on further CPS domains to investigate how SDC-Scissor performs when safety criteria concern new types of safety-critical faults, different from those inves-tigated in this study. Finally, we want to investigate different meta-heuristics to enable test case generation based on the designed feature sets.

Acknowledgment

Sebastiano Panichella, Sajad Khatiri, and Christian Birchler gratefully acknowledges the Horizon 2020 (EU Commission) support for the project COS-MOS (DevOps for Complex Cyber-physical Systems), Project No. 957254-COSMOS).

References

1. Nvidia drive constellation (2020). URL https://developer.nvidia.com/drive/ drive-constellation
2. Abdessalem, R.B., Nejati, S., Briand, L.C., Stifter, T.: Testing vision-based control systems using learnable evolutionary algorithms. In: Proceedings of the 40th International Conference on Software Engineering. ACM (2018). DOI 10.1145/3180155.3180160. URL https://doi.org/10.1145/3180155.3180160
3. Abdessalem, R.B., Panichella, A., Nejati, S., Briand, L.C., Stifter, T.: Testing au-tonomous cars for feature interaction failures using many-objective search. In: IEEE/ACM International Conference on Automated Software Engineering (ASE), pp. 143{154. IEEE (2018)
4. Academies of Sciences: A 21st Century Cyber-Physical Systems Education. National Academies Press (2017)
5. Afzal, A., Katz, D.S., Goues, C.L., Timperley, C.S.: A study on the challenges of using robotics simulators for testing (2020)
6. Arcuri, A., Briand, L.C.: A hitchhiker's guide to statistical tests for assessing randomized algorithms in software engineering. Softw. Test. Veri cation Reliab. 24(3), 219{250 (2014). DOI 10.1002/stvr.1486. URL https://doi.org/10.1002/stvr.1486
7. Arrieta, A., Wang, S., Arruabarrena, A., Markiegi, U., Sagardui, G., Etxeberria, L.: Multi-objective black-box test case selection for cost-e ectively testing simulation mod-els. In: Proceedings of the Genetic and Evolutionary Computation Conference, pp. 1411{1418 (2018)
8. Arrieta, A., Wang, S., Markiegi, U., Sagardui, G., Etxeberria, L.: Employing multi-objective search to enhance reactive test case generation and prioritization for testing in-dustrial cyber-physical systems. IEEE Trans. Ind. Informatics 14(3), 1055{1066 (2018). DOI 10.1109/TII.2017.2788019. URL https://doi.org/10.1109/TII.2017.2788019
9. Arrieta, A., Wang, S., Sagardui, G., Etxeberria, L.: Search-based test case selection of cyber-physical system product lines for simulation-based validation. In: H. Mei (ed.) Proceedings of the 20th International Systems and Software Product Line Conference, SPLC 2016, Beijing, China, September 16-23, 2016, pp. 297{306. ACM (2016). DOI 10.1145/2934466.2946046. URL https://doi.org/10.1145/2934466.2946046
10. Arrieta, A., Wang, S., Sagardui, G., Etxeberria, L.: Search-based test case prioritization for simulation-based testing of cyber-physical system product lines. J. Syst. Softw. 149, 1{34 (2019). DOI 10.1016/j.jss.2018.09.055. URL https://doi.org/10.1016/j.jss. 2018.09.055

11. Baeza-Yates, R., Ribeiro-Neto, B.A.: Modern Information Retrieval - the concepts and technology behind search, Second edition. Pearson Education Ltd., Harlow, England (2011). URL http://www.mir2ed.org/
12. Baheti, R., Gill, H.: Cyber-physical systems. The impact of control technology 12(1), 161{166 (2011)
13. BeamNG GmbH: BeamNG.tech. https://www.beamng.gmbh/research. Accessed: 2018-10-11
14. Bezerra, M.E.R., Oliveira, A.L.I., Meira, S.R.L.: A constructive rbf neural network for estimating the probability of defects in software modules. In: 2007 International Joint Conference on Neural Networks, pp. 2869{2874 (2007)

15. Bondi, E., Dey, D., Kapoor, A., Piavis, J., Shah, S., Fang, F., Dilkina, B., Hannaford, R., Iyer, A., Joppa, L., Tambe, M.: AirSim-w: A simulation environment for wildlife conservation with uavs. In: E.W. Zegura (ed.) Proceedings of the 1st ACM SIGCAS Conference on Computing and Sustainable Societies, COMPASS, pp. 40:1{40:12. ACM (2018). DOI 10.1145/3209811.3209880. URL https://doi.org/10.1145/3209811.3209880

16. Briand, L.C., Nejati, S., Sabetzadeh, M., Bianculli, D.: Testing the untestable: model testing of complex software-intensive systems. In: L.K. Dillon, W. Visser, L.A. Williams (eds.) Proceedings of the 38th International Conference on Software Engineering, ICSE 2016, Austin, TX, USA, May 14-22, 2016 - Companion Volume, pp. 789{792. ACM (2016). DOI 10.1145/2889160.2889212. URL https://doi.org/10.1145/2889160. 2889212

17. Canfora, G., Lucia, A.D., Penta, M.D., Oliveto, R., Panichella, A., Panichella, S.: Multi-objective cross-project defect prediction. In: Sixth IEEE International Conference on Software Testing, Veri cation and Validation, ICST 2013, Luxembourg, Luxembourg, March 18-22, 2013, pp. 252{261. IEEE Computer Society (2013). DOI 10.1109/ICST. 2013.38. URL https://doi.org/10.1109/ICST.2013.38

18. Caruana, R., Niculescu-mizil, A.: An empirical comparison of supervised learning algorithms. In: In Proc. 23 rd Intl. Conf. Machine learning (ICML'06, pp. 161{168 (2006)

19. Ceylan, E., Kutlubay, F.O., Bener, A.B.: Software defect identi cation using machine learning techniques. In: 32nd EUROMICRO Conference on Software Engineering and Advanced Applications (EUROMICRO'06), pp. 240{247 (2006)

20. Chen, H.: Applications of cyber-physical system: A literature review. Journal of Indus-trial Integration and Management 02(03), 1750012 (2017)

21. Chen, T.Y., Lau, M.F.: Dividing strategies for the optimization of a test suite. Infor-mation Processing Letters 60(3), 135{141 (1996)

22. CNX, O.: Openstax university physics (2021). URL http://cnx.org/contents/ d50f6e32-0fda-46ef-a362-9bd36ca7c97d@10.16

23. CollabNet: Anti-patterns in the continuous delivery (cd) practice (2014). URL https://goo.gl/JVxeBr

24. Dalboni, M., Soldati, A.: Soft-body modeling: A scalable and e cient formulation for control-oriented simulation of electric vehicles. In: IEEE Transportation Electri cation Conference and Expo (ITEC), pp. 1{6 (2019)

25. Dosovitskiy, A., Ros, G., Codevilla, F., Lopez, A.M., Koltun, V.: CARLA: an open urban driving simulator. In: 1st Annual Conference on Robot Learning, CoRL 2017, Proceedings of Machine Learning Research, vol. 78, pp. 1{16. PMLR (2017). URL http://proceedings.mlr.press/v78/dosovitskiy17a.html

26. Electronics, L.: Lgsvl simulator. URL https://www.lgsvlsimulator.com/

27. Flores-Garc a, E., Kim, G., Yang, J., Wiktorsson, M., Noh, S.D.: Analyzing the char-acteristics of digital twin and discrete event simulation in cyber physical systems. In: B. Lalic, V.D. Majstorovic, U. Marjanovic, G. von Cieminski, D. Romero (eds.) Ad-vances in Production Management Systems. Towards Smart and Digital Manufactur-ing - IFIP WG 5.7 International Conference, APMS 2020, Novi Sad, Serbia, August 30 - September 3, 2020, Proceedings, Part II, IFIP Advances in Information and Communication Technology, vol. 592, pp. 238{244. Springer (2020). DOI 10.1007/ 978-3-030-57997-5n 28. URL https://doi.org/10.1007/978-3-030-57997-5_28

28. Frank, E., Hall, M.A., Holmes, G., Kirkby, R., Pfahringer, B., Witten, I.H.: Weka: A machine learning workbench for data mining., pp. 1305{1314. Springer, Berlin (2005). URL http://researchcommons.waikato.ac.nz/handle/10289/1497

29. Gambi, A., Huynh, T., Fraser, G.: Generating e ective test cases for self-driving cars from police reports. In: M. Dumas, D. Pfahl, S. Apel, A. Russo (eds.) Proceedings of the ACM Joint Meeting on European Software Engineering Conference and Symposium on the Foundations of Software Engineering, ESEC/SIGSOFT FSE 2019, Tallinn, Estonia, August 26-30, 2019, pp. 257{267. ACM (2019). DOI 10.1145/3338906.3338942. URL https://doi.org/10.1145/3338906.3338942

30. Gambi, A., Huynh, T., Fraser, G.: Generating e ective test cases for self-driving cars from police reports. In: Proceedings of the 2019 27th ACM Joint Meeting on European Software Engineering Conference and Symposium on the Foundations of Software En-gineering - ESEC/FSE 2019. ACM Press (2019). DOI 10.1145/3338906.3338942. URL https://doi.org/10.1145/3338906.3338942

31. Gambi, A., Mueller, M., Fraser, G.: AsFault: Testing self-driving car software using search-based procedural content generation. In: 2019 IEEE/ACM 41st International Conference on Software Engineering: Companion Proceedings (ICSE-Companion). IEEE (2019). DOI 10.1109/icse-companion.2019.00030. URL https://doi.org/10. 1109/icse-companion.2019.00030

32. Gambi, A., Muller, M., Fraser, G.: Automatically testing self-driving cars with search-based procedural content generation. In: D. Zhang, A. M ller (eds.) Proceedings of the 28th ACM SIGSOFT International Symposium on Software Testing and Analysis, ISSTA 2019, Beijing, China, July 15-19, 2019, pp. 318{328. ACM (2019). DOI 10.1145/3293882.3330566. URL https://doi.org/10.1145/3293882.3330566

33. Gonzalez, C.A., Varmazyar, M., Nejati, S., Briand, L.C., Isasi, Y.: Enabling model testing of cyber-physical systems. In: Proceedings of the 21th ACM/IEEE International Conference on Model Driven Engineering Languages and Systems, MODELS '18, p. 176{186. Association for Computing Machinery, New York, NY, USA (2018). DOI 10.1145/3239372.3239409. URL https://doi.org/10.1145/3239372.3239409

34. Guardian, T.: Self-driving uber kills arizona woman in rst fatal crash involving pedestrian (2018). URL https://www.theguardian.com/technology/2018/mar/19/ uber-self-driving-car-kills-woman-arizona-tempe

35. Hemon, A., Lyonnet, B., Rowe, F., Fitzgerald, B.: From agile to DevOps: Smart skills and collaborations. Information Systems Frontiers 22(4), 927{945 (2019). DOI 10.1007/s10796-019-09905-1. URL https://doi.org/10.1007/s10796-019-09905-1

36. Ho, T.K.: The random subspace method for constructing decision forests. IEEE Transactions on Pattern Analysis and Machine Intelligence 20(8), 832{844 (1998). DOI 10.1109/34.709601. URL https://doi.org/10.1109/34.709601

37. Ingrand, F.: Recent trends in formal validation and veri cation of autonomous robots software. In: 3rd IEEE International Conference on Robotic Computing, IRC 2019, Naples, Italy, February 25-27, 2019, pp. 321{328 (2019)

38. International, T.: Simcenter prescan (2020). URL https://tass.plm.automation. siemens.com/prescan

39. Kalra, N., Paddock, S.: Driving to safety: How many miles of driving would it take to demonstrate autonomous vehicle reliability? Transportation Research Part A: Policy and Practice 94, 182{193 (2016). DOI 10.1016/j.tra.2016.09.010

40. Kaur, A., Malhotra, R.: Application of random forest in predicting fault-prone classes. In: 2008 International Conference on Advanced Computer Theory and Engineering, pp. 37{43 (2008)

41. Khatiri, S., Birchler, C., Bosshard, B., Gambi, A., Panichella, S.: Machine Learning-based Test Selection for Simulation-based Testing of Self-driving Cars Software" (2021). DOI 10.5281/zenodo.5085252. URL https://zenodo.org/record/5085252# .YOhRYJMzb7E

42. Kim, J., Chon, S., Park, J.: Suggestion of testing method for industrial level cyber-physical system in complex environment. In: 2019 IEEE International Conference on Software Testing, Veri cation and Validation Workshops (ICSTW). IEEE (2019). DOI 10.1109/icstw.2019.00043. URL https://doi.org/10.1109/icstw.2019.00043

43. Li, G., Li, Y., Jha, S., Tsai, T., Sullivan, M., Hari, S.K.S., Kalbarczyk, Z., Iyer, R.: Av-fuzzer: Finding safety violations in autonomous driving systems. In: 2020 IEEE 31st International Symposium on Software Reliability Engineering (ISSRE), pp. 25{36. IEEE (2020)

44. Ling, C.X., Li, C.: Data mining for direct marketing: Problems and solutions. In: Proceedings of the Fourth International Conference on Knowledge Discovery and Data Mining, KDD'98, p. 73{79. AAAI Press (1998)

45. Loquercio, A., Kaufmann, E., Ranftl, R., Dosovitskiy, A., Koltun, V., Scaramuzza, D.: Deep drone racing: From simulation to reality with domain randomization. IEEE Transactions on Robotics 36(1), 1{14 (2020)

46. Matinnejad, R., Nejati, S., Briand, L., Bruckmann, T., Poull, C.: Automated model-in-the-loop testing of continuous controllers using search. In: International Symposium on Search Based Software Engineering, pp. 141{157. Springer (2013)

47. Mesit, J., Guha, R.K.: A general model for soft body simulation in motion. In: S. Jain, R.R.C. Jr., J. Himmelspach, K.P. White, M.C.F. 0001 (eds.) Winter Simulation Conference, pp. 2690{2702. IEEE (2011). URL http://dl.acm.org/citation.cfm?id=2431518

48. Nucci, D.D., Panichella, A., Zaidman, A., Lucia, A.D.: A test case prioritization genetic algorithm guided by the hypervolume indicator. IEEE Trans. Software Eng. 46(6), 674{696 (2020). DOI 10.1109/TSE.2018.2868082. URL https://doi.org/10.1109/ TSE.2018.2868082

49. Panichella, S., Di Sorbo, A., Guzman, E., Visaggio, C.A., Canfora, G., Gall, H.C.: How can I improve my app? Classifying user reviews for software maintenance and evolution. In: Proc. Int'l Conf. on Software Maintenance and Evolution (ICSME), pp. 281{290 (2015)

50. Panichella, S., Gambi, A., Zampetti, F., Riccio, V.: Sbst tool competition 2021. In: International Conference on Software Engineering, Workshops, Madrid, Spain, 2021. ACM (2021)

51. Refaeilzadeh, P., Tang, L., Liu, H.: Cross-Validation, pp. 532{538. Springer US, Boston, MA (2009). DOI 10.1007/978-0-387-39940-9 565. URL https://doi.org/10.1007/ 978-0-387-39940-9_565

52. Riccio, V., Tonella, P.: Model-based exploration of the frontier of behaviours for deep learning system testing. In: Proceedings of the ACM Joint European Software Engineering Conference and Symposium on the Foundations of Software Engineering, ESEC/FSE '20, p. 13 pages. Association for Computing Machinery (2020). DOI 10.1145/3368089.3409730

53. Rothermel, G., Harrold, M.J., Ostrin, J., Hong, C.: An empirical study of the effects of minimization on the fault detection capabilities of test suites. In: Proceedings of the International Conference on Software Maintenance, pp. 34{44. IEEE CS Press (1998)

54. Rothermel, G., Untch, R., Chu, C., Harrold, M.: Test case prioritization: an empirical study. In: Software Maintenance, 1999. (ICSM '99) Proceedings. IEEE International Conference on, pp. 179{188 (1999). DOI 10.1109/ICSM.1999.792604

55. Shin, S.Y., Chaouch, K., Nejati, S., Sabetzadeh, M., Briand, L.C., Zimmer, F.: Hitecs: A uml profile and analysis framework for hardware-in-the-loop testing of cyber physical systems. In: Proc. Int'l Conf. on Model Driven Engineering Languages and Systems (MODELS), pp. 357{367. ACM (2018)

56. Shin, S.Y., Nejati, S., Sabetzadeh, M., Briand, L.C., Zimmer, F.: Test case prioritization for acceptance testing of cyber physical systems: a multi-objective search-based approach. In: F. Tip, E. Bodden (eds.) Proceedings of the 27th ACM SIGSOFT International Symposium on Software Testing and Analysis, ISSTA 2018, Amsterdam, The Netherlands, July 16-21, 2018, pp. 49{60. ACM (2018). DOI 10.1145/3213846.3213852. URL https://doi.org/10.1145/3213846.3213852

57. Shokry, H., Hinchey, M.: Model-based verification of embedded software. Computer 42(4), 53{59 (2009). DOI 10.1109/MC.2009.125

58. Sontges, S., Althoff, M.: Computing the drivable area of autonomous road vehicles in dynamic road scenes. IEEE Trans. Intell. Transp. Syst. 19(6), 1855{1866 (2018). DOI 10.1109/TITS.2017.2742141. URL https://doi.org/10.1109/TITS.2017.2742141

59. The-Washington-Post: Uber's radar detected elaine herzberg nearly 6 seconds before she was fatally struck, but \the system design did not include a consideration for jaywalking pedestrians" so it didn't react as if she were a person. (2019). URL https://mobile. twitter.com/faizsays/status/1191885955088519168

60. Tolles, J., Meurer, W.J.: Logistic regression. JAMA 316(5), 533 (2016). DOI 10.1001/ jama.2016.7653. URL https://doi.org/10.1001/jama.2016.7653

61. Torngren, M., Sellgren, U.: Complexity Challenges in Development of Cyber-Physical Systems, pp. 478{503. Springer International Publishing, Cham (2018)

62. Xu, J., Luo, Q., Xu, K., Xiao, X., Yu, S., Hu, J., Miao, J., Wang, J.: An automated learning-based procedure for large-scale vehicle dynamics modeling on baidu apollo platform. In: 2019 IEEE/RSJ International Conference on Intelligent Robots and Sys-tems, IROS, pp. 5049{5056. IEEE (2019). DOI 10.1109/IROS40897.2019.8968102. URL https://doi.org/10.1109/IROS40897.2019.8968102

63. Yohanandhan, R.V., Elavarasan, R.M., Manoharan, P., Mihet-Popa, L.: Cyber-physical power system (CPPS): A review on modeling, simulation, and analysis with cyber secu-rity applications. IEEE Access 8, 151019{151064 (2020). DOI 10.1109/ACCESS.2020. 3016826. URL https://doi.org/10.1109/ACCESS.2020.3016826

64. Yoo, S., Harman, M.: Using hybrid algorithm for Pareto efficient multi-objective test suite minimisation. Journal of Systems and Software 83(4), 689{701 (2010)

65. Yoo, S., Harman, M.: Regression testing minimization, selection and prioritization: a survey. Software Testing, Veri cation and Reliability 22(2), 67{120 (2012)
66. Zampetti, F., Vassallo, C., Panichella, S., Canfora, G., Gall, H., Di Penta, M.: An empirical characterization of bad practices in continuous integration. In: Empirical Software Engineering (2019)
67. Zapridou, E., Bartocci, E., Katsaros, P.: Runtime veri cation of autonomous driving systems in carla. In: J. Deshmukh, D. Nickovi (eds.) Runtime Veri cation, pp. 172{ 183. Springer International Publishing, Cham (2020)